\newlength{\absize}
\documentclass[11pt]{article}
\usepackage{graphicx}
\usepackage{latexsym}
\usepackage{amssymb}
\usepackage{bbm}
\usepackage{setspace}
\usepackage{braket}
\renewcommand{\arraystretch}{2}

\newcommand{\figsize}{\small}

\input prepictex
\input pictex
\input postpictex
\newdimen\tdim
\tdim=\unitlength
\def\stpltsmbl{\setplotsymbol ({\small .})}

\newbox\sru
\setbox\sru=\hbox{\beginpicture
\setcoordinatesystem units <\tdim,\tdim>
\stpltsmbl
\setquadratic
\plot
  0.0   0.0
  4.8   1.5
  7.5   5.0
  7.3   8.5
  5.0  10.0
  2.7   8.5
  2.5   5.0
  5.2   1.5
 10.0   0.0
/
\endpicture}
\def\springru #1 #2 *#3 /{\multiput {\copy\sru}  at
#1 #2 *#3 10 0 /}

\renewcommand{\bar}{\overline}

\newcommand{\cO}{\mathcal{O}}
\newcommand{\cB}{\mathcal{B}}
\newcommand{\cC}{\mathcal{C}}
\newcommand{\cD}{\mathcal{D}}

\newcommand{\be}{\begin{equation}}
\newcommand{\ee}{\end{equation}}
\newcommand{\bea}{\begin{eqnarray}}
\newcommand{\eea}{\end{eqnarray}}

\newcommand{\comment}[1]{}

\begin{document}

\thispagestyle{empty}
\pagestyle{empty}
\newcommand{\starttext}{\newpage\normalsize
 \pagestyle{plain}
 \setlength{\baselineskip}{3ex}\par
 \setcounter{footnote}{0}
 \renewcommand{\thefootnote}{\arabic{footnote}}
 }
\newcommand{\preprint}[1]{\begin{flushright}
 \setlength{\baselineskip}{3ex}#1\end{flushright}}
\renewcommand{\title}[1]{\begin{center}\LARGE
 #1\end{center}\par}
\renewcommand{\author}[1]{\vspace{2ex}{\large\begin{center}
 \setlength{\baselineskip}{3ex}#1\par\end{center}}}
\renewcommand{\thanks}[1]{\footnote{#1}}
\renewcommand{\abstract}[1]{\vspace{2ex}\normalsize\begin{center}
 \centerline{\bf Abstract}\par\vspace{2ex}\parbox{\absize}{#1
 \setlength{\baselineskip}{2.5ex}\par}
 \end{center}}

%\preprint{}
\title{Non-perturbative Effects and Unparticle Physics 
in Generalized Schwinger Models}
\author{
 Howard~Georgi\thanks{\noindent \tt hgeorgi@fas.harvard.edu}
and
Bea~Noether\thanks{\noindent \tt bea\_noether@berkeley.edu} 
\\ \medskip
Center for the Fundamental Laws of Nature\\
Jefferson Physical Laboratory \\
Harvard University \\
Cambridge, MA 02138\\
 }
\date{\today}
\abstract{

We analyze generalizations of the Schwinger model with more massless
fermions and more vector fields.
We focus on models with the gauge structure
of ``diagonal color $SU(n)$'' but
unlike previous investigators, we do not assume that all the gauge boson masses
are the same. 
Unlike the Schwinger model,  
these are Banks-Zaks models with
conformal sectors that survive 
at long distances.  In addition to local operators that go to
``unparticle operators'' with non-zero anomalous dimensions at long
distances, they contain local operators like the
$\bar\psi_L\psi_R$ operator in the Schwinger model which go to constants at
long distances.  These operators have calculable vacuum 
expectation values (up to phases).  Cluster decomposition applied to
correlation functions involving these operators yields nontrivial and
calculable non-perturbative constraints on correlation functions.  One
consequence is ``conformal coalescence'' in
which linear combinations of short distance operators disappear from
the long-distance theory, leaving only one kind of unparticle stuff in the
low-energy theory.  
We believe that our detailed analysis of diagonal color $SU(n)$
paints an appealing picture of unparticle operators as the result of an
incomplete binding of the massless fermions.  We complete the picture (and
the binding) by
analyzing the diagonal color $U(n)$ model with a very small $U(1)$ coupling and
thus a gauge boson with a dynamical mass much smaller than the other masses
in the model.  This model has a mass gap and we can see explicitly the
transition from 
free-fermion behavior at short distances to unparticle physics at
intermediate distances to the physics of massive particles at long
distances.
}

%\newpage
%\tableofcontents
\starttext

\section{Introduction\label{sec-intro}}
The Schwinger model~\cite{Schwinger:1962tp} of the electrodynamics of a
massless fermion in 1+1 dimensions is a completely solvable quantum field
theory that exhibits some of the features of QCD:
confinement~\cite{Wilson:1974sk, Coleman:1975pw}\footnote{There is a
literature on confinement and the slightly different effect of screening in
1+1 dimension.  See \cite{Gross:1995bp} and references therein.}; 
an anomalous chiral $U(1)$\cite{Nielsen:1976hs}; 
instantons and $\theta$ vacua~\cite{Maiella:1977gn, Lee:1979sm,
Adam:1993fc, Smilga:1994hc, Radozycki:1998xs};  
non-zero vacuum expectation values (VEVs) of composite
operators~\cite{Smilga:1992hx,Jayewardena:1988td,Hetrick:1988yg}. 
In this note, we consider generalizations of the Schwinger model with more massless
fermions and gauge fields.  These models can be solved explicitly
if all the gauge couplings commute.  We developed the tools to study these
models in 
\cite{Georgi:2019pje} where we analyzed the most general Sommerfield
models~\cite{Sommerfield:1964} from which we can construct the generalized
Schwinger models.

We focus on a generalization that has been studied by other authors ---
the diagonal color $SU(n)$ models with a single ``flavor'' but $n$-colors of
massless fermions coupled to $n$$-$$1$ gauge bosons with the diagonal
$SU(n)$ charges.~\cite{Belvedere:1978fj,Steinhardt:1980rz,
GamboaSaravi:1981zd, Belvedere:1986ik}
Unlike previous works, we focus on these models
as examples of Banks-Zaks
models~\cite{Banks:1981nn} with weak coupling at short distances and
a conformal sector with non-trivial anomalous dimensions at long
distances.  We calculate the correlation functions of low-dimension 
gauge-invariant operators
to study the transition from short
distance behavior dominated by nearly free fermions to long distance
behavior dominated by the conformal sector. 
Our models are also new because 
we do not assume
that the gauge couplings are equal.
We therefore have a spectrum of
boson masses that makes the transition much more complicated and we believe
more interesting.  

Like the Schwinger model, diagonal $SU(n)$ color models have
composite operators that behave like constants at long distances. 
Cluster decomposition requires that these 
have non-zero vacuum expectation values, and the magnitudes of these VEVs
are exactly calculable~\cite{Smilga:1992hx}.  These VEVs break symmetries of the
classical Lagrangian that are broken in the quantum theory due to
instanton effects which do not show up to any order in
perturbation theory, unlike the properties of the model
that can be obtained by summing the perturbation series.\footnote{Thus we
will refer to the VEVs and the related phenomena that we discuss below as
``non-perturbative.''}   

For large $n$, diagonal $SU(n)$ color models have exponentially many
short-distance operators that flow to low-dimension operators at long
distances, with anomalous dimensions that go to zero like $1/n$. 
We use cluster decomposition along with perturbative calculations
to calculate non-perturbative contributions to
correlators of these low-dimension operators,  and we observe a phenomenon
we call ``conformal coalescence'' in which only one linear combination with
each dimension and chiral charge survives in the long-distance theory.

In section~\ref{sec-zdops}, we review the Schwinger model, emphasizing the
operators that go to constants at long distances, which we call
``zero-dimension operators'' (ZDOPs). 
In section~\ref{sec-color}, we define diagonal color $SU(n)$ models. We
identify the local operators with anomalous dimension less than 1 at long
distances. These comprise the ZDOPs and ``low-dimension operators'' (LDOPs)
with non-zero anomalous dimensions at long distances.
In section~\ref{sec-masses}
we write down the perturbative 2-point functions for ZDOPs and LDOPs for
arbitrary gauge boson masses and illustrate the effect of multiple masses
by plotting some representative correlators for
space-like separations.  We also discuss the dependence of ZDOP VEVs on
different gauge boson masses and write down an explicit example in
appendix~\ref{sec-vevmasses}. 
We further note that while we
discuss diagonal color because the coupling structure is familiar,
our analysis applies to a large class of couplings that give the
same ZDOPs, LDOPs and anomalous dimensions
but have different dependence on the
gauge boson masses.  
In section~\ref{sec-3pt} and appendix~\ref{sec-signs}
we write down the perturbative 3-point functions involving ZDOPs and LDOPs 
and use the result with cluster decomposition
to find relations among the phases of the ZDOP VEVs.
In section~\ref{sec-rpoint} we extend the analysis of
the previous section to arbitrary perturbative correlators of ZDOPs and LDOPs.
In section~\ref{sec-coalescence} we apply the results of the previous
sections and cluster decomposition
to calculate all perturbative and non-perturbative correlators. We show
that up to normalization factors (which we calculate) the perturbative and
non-perturbative correlators depend only on the chiral charges of the
operators.  We interpret this to mean 
that all normalized operators with the same chiral
charge flow to the same operator in the low-energy theory.  This is what we
call conformal coalescence.
In section~\ref{sec-ib} we suggest a way of thinking about the physics in
these models.  The ZDOPs are associated with complete binding of the
massless fermions into massive particle states.  The LDOPs, unparticle
operators, are the result of 
``incomplete binding'' in which the massless fermions are neither free nor
bound into particles. We explore this picture further by adding a small
$U(1)$ gauge 
coupling to the model.  Then all the fermions are bound into particles, but
with radically different masses.
Section~\ref{sec-end} contains some suggestions for further exploration.

\section{ZDOPs\label{sec-zdops}}

To think about generalizations of the Schwinger model, it is useful
to consider them as limits of Sommerfield models, with massless
fermions with commuting couplings to one or more vector bosons, but with
bare masses for the 
vector bosons~\cite{Georgi:2019pje}.  All such Sommerfield models are
examples of Bank-Zaks 
models~\cite{Banks:1981nn} with weak coupling at short distances and
conformal invariance with non-trivial anomalous dimensions at long
distances.  We will be interested not just in the properties of the
long-distance conformal theory, but in the sometimes complicated transition
from short distance to long distance.  We define what we mean by 
``generalized Schwinger models'' as the limits of the
gauge-invariant sectors of Sommerfield models as the vector boson masses go
to zero.  

We the use the results of \cite{Georgi:2019pje} for gauge-invariant
operators in models with gauge bosons $A^\mu_j$ for $j=1$ to
$n_A$ with diagonal vector couplings $e_{j\alpha}$ to fermions
$\psi_\alpha$, for $\alpha=1$ to $n_F\geq n_A$ where
\begin{equation}
m_j^2\,\delta_{jk}=\sum_\alpha\frac{e_{j\alpha}\,e_{k\alpha}}{\pi}
\label{m02}
\end{equation}
This describes the limit of the gauge-invariant sector of a Sommerfield
model in which all the vector boson mass terms in the Lagrangian vanish and
the entire theory is gauge-invariant.  We call this the Schwinger
point.~\cite{Georgi:2019vyk} 
While any such model with commuting couplings can be solved
explicitly, we can always make a unitary
transformation of the fermion fields and an orthogonal transformation of
the gauge fields to get the form (\ref{m02}).  This form is very convenient
both because the couplings to the fermions are diagonal and because
the $m_j$s are the
physical masses of the bosons associated with the gauge fields,
$A^\mu_j$.  Thus we always assume that we have done the necessary
transformations to obtain (\ref{m02}).

Then in gauge-invariant combinations of
LH fermions $\psi_{\alpha\,1}$ and RH fermions
$\psi_{\alpha\,2}$, the Lagrangian fields can be replaced by products of
free fields 
\be
\psi_{\alpha\,1} 
\to e^{-i\bigl(\sum_j(e_{j\alpha}/m_j)([\cC]_j-[\cB]_j)\bigr)}\Psi_{\alpha\,1}
\quad\quad
\psi_{\alpha\,2} 
\to
e^{i\bigl(\sum_j(e_{j\alpha}/m_j)([\cC]_j-[\cB]_j)\bigr)}\Psi_{\alpha\,2}
\label{psi-alpha}
\ee
$\Psi$s are free massless fermion fields, $\cB_j$ are pseudo-scalar fields
with mass $m_j$, and $\cC_j$ are massless ghosts.
The long-distance anomalous dimension of a gauge-invariant operator comes
from the 
free fermion propagators and the exponential of the ghost field
propagators.

We consider gauge-invariant operators that have at most
one power of each type of fermion field (LH and RH
count as different types but we do not allow $\psi$ and $\psi^*$ of the same
type and handedness).  
Such operators appear as the first term in an operator product expansion (to
all orders in perturbation theory) of
the product of the component fields with no necessity for normal ordering
and since the short distance physics is free, we can bring the fermion
fields to the same point without divergences.
But more importantly, these operators have the
lowest dimensions at long distances.\footnote{Of course, this does not mean
that other operators are uninteresting.  In particular, the Noether
currents and the energy momentum tensor are not in this class.} 

In the classic Schwinger model, $n_A=n_F=1$.  There is only one kind of
gauge boson and fermion, so we can drop the $j$ and $\alpha$.  Then the local
operators with no $\psi^*\psi$ pairs look like
\begin{equation}
\cO_{12}=\psi_1^*(x)\,\psi_2(x)
\quad\mbox{and}\quad
\cO^*_{12}=\psi_2^*(x)\,\psi_1(x)
\label{upops}
\end{equation}

To compute the dimension of the operator $\cO_{12}$ 
in the long-distance theory, we
calculate the 2-point function of the operator and its complex conjugate,
which for this class of operators is
{\renewcommand{\arraystretch}{2.5}
\begin{equation}
\begin{array}{c}
\displaystyle
\braket{0|T\,\cO_{12}(x)\,\cO^*_{12 }(0)|0}
\\ \displaystyle
=S_1(x)\,
S_2(x)
\exp\left[\frac{2e^2}{\pi m^2}
\left[K_0\left(m\sqrt{-x^2 + i\epsilon}\right) + \ln\left(\xi m\sqrt{-x^2 +
i\epsilon}\right)\right]\right]
\\ \displaystyle
=\frac{(\xi
m)^{2(e^2/\pi)/m^2}}{(2\pi)^2}\,
\exp\left[\frac{2e^2}{\pi m^2}
\left[K_0\left(m\sqrt{-x^2 + i\epsilon}\right) \right]\right]
\left(\frac{1}{-x^2+i\epsilon}\right)^{1-(e^2/\pi)/m^2}
\label{d1/2-2pt}
\end{array}
\label{oostar}
\end{equation}
}%
where
\begin{equation}
\xi\equiv e^{\gamma_E}/2
\label{xi}
\end{equation}
At long distances, the exponential goes to one and we can read off the
dimension of the operator as
\begin{equation}
1-(e^2/\pi)/ m^2
\label{sdim}
\end{equation}
Because of (\ref{m02}), this vanishes.
Thus the Schwinger model
has a mass gap and there is no conformal sector.
In the operator language, what is happening is that the effect of the ghost
propagator in the exponential is to decrease the dimension from its
free-field value.  For this operator in the
Schwinger model, the ghost exactly cancels the free field behavior from the
fermion propagators and the 2-point function goes to a constant at long
distances.  
Then 
cluster decomposition implies that the operators
must have non-zero vacuum expectation values, because it must be that
\begin{equation}
\braket{0|{\rm T}\,\cO_{12} (x)\,\cO_{12}^* (0)|0}
\mathop{\longrightarrow}\limits_{-x^2\to\infty}
\braket{0|\cO_{12} (0)|0}\,\braket{0|\cO_{12}^* (0)|0}
\label{2pt-local-cluster}
\end{equation}
This means the vacuum at the Schwinger point must be degenerate with
\begin{equation}
\braket{0|\cO_{12} (0)|0}
=\frac{\xi m}{2\pi}\,e^{i\theta}
\quad\quad
\braket{0|\cO_{12}^* (0)|0}
=\frac{\xi m}{2\pi}\,e^{-i\theta}
\label{2pt-local-vev}
\end{equation}
where $\theta$ is the parameter that labels the different but physically
equivalent vacuum 
states, related by chiral transformations on the massless fermion
fields~\cite{Smilga:1992hx,Jayewardena:1988td,Hetrick:1988yg}.  

In ``diagonal color models''~\cite{Belvedere:1978fj,
Steinhardt:1980rz, GamboaSaravi:1981zd} we find many examples of 
long-distance 0-dimension operators, which we call Zero-Dimension
Operators or ZDOPs.
Because the smallest dimension in the long-distance conformal theory
consistent with unitarity is 
$0$~\cite{Mack:1975je}, ZDOPs can
only occur at a boundary of parameter space, like the Schwinger point. They
do not occur in generalized Sommerfield models.
We will see that the ZDOPs are associated with non-perturbative
effects that have no analog in the original Schwinger model.

Before describing diagonal color 
models with ZDOPs, we should note that the existence of
ZDOPs is not generic.  It requires special relationships among couplings.
A simple example without ZDOPs is a model with two massless
fermions with couplings $e$ and $2e$ to one gauge boson.  At the Schwinger point
there are non-derivative composite operators with long-distance dimensions
$1/5$, $4/5$, 
and greater,
but no ZDOPs!

\section{Diagonal color\label{sec-color}}

Because we wish to focus on models that are solvable
in the same sense as the Schwinger model, we will consider only models with
diagonal gauge couplings, and thus $SU(n_c)$ color is not available to us.  
However, a related class of
models with multiple gauge bosons has been studied in the literature under
the name ``diagonal color''~\cite{Steinhardt:1980rz}.
In an $SU(n_c)$
diagonal color model, there are $n_c$ colors (and one flavor) of fermions and
$n_c$$-$$1$ gauge bosons with couplings proportional to the diagonal traceless 
$n_c$$\times$$n_c$ matrices in the conventional basis
\be
[T^D_j]_{\alpha\beta}=e_{j\alpha}\,\delta_{\alpha\beta}
%\label{}
\ee
where
{\renewcommand{\arraystretch}{1.1}
\begin{equation}
e_{j \alpha}=\frac{e_j}{\sqrt{2j (j +1)}}\times\left\{
\begin{array}{l}
1\mbox{~~for $\alpha\leq j $}\\
-j \mbox{~~for $\alpha= j +1$}\\
0\mbox{~~for $j +1<\alpha$}
\end{array}\right.
\mbox{~~where~~}e_j=\sqrt{2\pi}\,m_j
\label{e-diagonal-color}
\end{equation}}%
where $m_j$ are the physical boson masses.
We have chosen the normalizations in (\ref{e-diagonal-color}) so that
$e\,e^T/\pi$ is the diagonal 
mass-squared matrix of the massive bosons, as in (\ref{m02})
so we are at the Schwinger point.  

We cannot make arbitrary color transformations on the fermion fields
because they do not leave the $T^D_j$ diagonal.
If the $m_j$ (and thus the $e_j$s in
(\ref{e-diagonal-color})) are all the same, we can make discrete
color transformations that permute the fermion fields because these can be
compensated by an orthogonal transformation on the gauge fields.  But in
general, if the gauge boson masses are different, we cannot make any
non-Abelian transformations without changing the short-distance physics.

At the classical level, in addition to the discrete symmetries,
the diagonal $SU(n)$ color Lagrangian has a
$U(1)^{2n}$ chiral symmetry on the fermion fields, 
with $n$ scalar charges and
$n$ axial charges.  $n$$-$$1$ of the scalar charges are associated
with the gauge symmetries.  These are
irrelevant to us because we only look at gauge singlets.  $n$$-$$1$ of the axial
charges are broken in the quantum theory by the gauge anomalies and
instanton effects. But this leaves two unbroken
$U(1)$s, associated with fermion number and chiral fermion number, or
equivalently left-fermion number and right-fermion number that are separately
conserved. 

In general, the physics depends on the masses and (related) couplings of the
different bosons but the dependence has a very special form because of (\ref{m02}).
When we put (\ref{e-diagonal-color}) into the expression for our fermions
in terms of free 
fields, (\ref{psi-alpha}), and calculate correlators, the couplings appear
only in the dimensionless combinations
\be
\lambda_{j\gamma_1\gamma_2}\equiv 
\frac{e_{j{\gamma_1}}e_{j{\gamma_2}}}{\pi m_j}
=\delta_{{\gamma_1},j+1}\delta_{{\gamma_2},j+1}
-\frac{1}{j+1}u^{j+1}_{\gamma_1} u^{j+1}_{\gamma_2}
+\frac{1}{j}u^{j}_{\gamma_1} u^{j}_{\gamma_2}
\label{ejejgeneral}
\ee
where
\be
u^j_\gamma=\left\{
\begin{array}{l}
\mbox{$1$ for $\gamma\leq j$}\\
\mbox{$0$ for $\gamma> j$}
\end{array}
\right.
\label{uj}
\ee
and the long-distance anomalous
dimensions 
depend only on the sum
\be
\lambda^n_{\gamma_1\gamma_2}\equiv \sum_{j=1}^{n-1}\lambda_{j\gamma_1\gamma_2}
%\label{}
\ee
Furthermore, $\lambda^n_{\gamma_1\gamma_2}$ has a very simple form
\begin{equation}
\lambda^n_{\gamma_1\gamma_2}=\delta_{\gamma_1\gamma_2}
-\frac{1}{n}u^n_{\gamma_1} u^n_{\gamma_2}
\label{eoverm}
\end{equation}

We will see that there are ZDOPs in $SU(n)$ diagonal color models.  But the
model also contains operators, which we will call Low-Dimension 
Operators (LDOPs), with anomalous dimensions that go to zero like $1/n$.  We
will be particularly interested in the LDOPs as well as the ZDOPs.
The ZDOPs and LDOPs all have zero fermion number.  

We can use (\ref{eoverm}) to identify the ZDOPs and LDOPs and find the anomalous
dimensions of the LDOPs.  For example consider the correlator
\be
\Braket{0|{\rm T}\,\psi^*_{\alpha 1}\psi_{\alpha 2}(x)
\psi^*_{\alpha 2}\psi_{\alpha 1}(0)|0}
%\label{}
\ee
Exactly the same calculation as in the classic Schwinger model shows that
the $\delta_{\gamma_1\gamma_2}$ term 
in (\ref{eoverm}) in the ghost coupling cancels the
free-fermion terms in the 2-point function leaving only the $1/n$ term.
Thus
$\psi^*_{\alpha 1}\psi_{\alpha 2}$
and $\psi^*_{\alpha 2}\psi_{\alpha 1}$
are LDOPs with 
dimension $1/n$.  

In fact, we can use (\ref{eoverm}) to compute 
the anomalous dimensions of all the LDOPs and identify the ZDOPs.
Define the gauge-invariant operator
\begin{equation}
\phi_{A;B}=
\phi_{\alpha_1\cdots \alpha_\mu;\beta_1\cdots \beta_\nu}\equiv 
\psi^*_{\alpha_11}\cdots\psi^*_{\alpha_\mu1}
\psi^*_{\beta_12}\cdots\psi^*_{\beta_\nu2}
\psi_{\beta_\nu1}\cdots\psi_{\beta_11}
\psi_{\alpha_\mu2}\cdots\psi_{\alpha_12}
\label{phi}
\end{equation}
where $A=\{\alpha_1\cdots\alpha_\mu\}$ and $B=\{\beta_1\cdots\beta_\nu\}$ are
sets of distinct fermion color indices 
with values from $1$ 
to $n$.  Because of Fermi statistics, 
no index can appear twice in $A$ or $B$.  It is important to 
note that $A$ and $B$ are sets,
not sequences. The order of the indices is irrelevant but once an order is
chosen for the indices of the $\psi^*$s, the {\bf opposite}\footnote{This
curious definition will be convenient because it eliminates many
uninteresting signs in correlators.} order must be used
for the indices in the $\psi$s. 
Then because the color indices in $A$ (for example) 
appear in the
opposite order in   $\psi^*_{\alpha_11}\cdots\psi^*_{\alpha_\mu1}$ and
$\psi_{\alpha_12}\cdots\psi_{\alpha_\mu2}$, the order of the color indices
in $A$ makes no difference. If we shuffle the order, anticommutation
introduces an even number of minus signs.  $B$ works the same way.  

The numbers $\mu$ and $\nu$ count the elements
\be
\mu\equiv \mathbf{N}(A)
\mbox{~and~}
\nu\equiv \mathbf{N}(B)
\label{muandnu}
\ee
$\phi$ carries global chiral charge $2(\mathbf{N}(A)-\mathbf{N}(B))$.  The
$\phi$s satisfy 
\begin{equation}
\phi^*_{A;B}
=
\phi_{B;A}
\label{phistar}
\end{equation}
We will be particularly
interested in $\phi$s for which there is no overlap
between the sets of indices
\be
A\cap B=\varnothing
\label{acapb}
\ee
because (as we will see) 
these will be the LDOPs and ZDOPs.
Using
(\ref{psi-alpha})
we can write $\phi$ in terms of free fields.
\be
{\renewcommand{\arraystretch}{1.7}\begin{array}{c}
\displaystyle
\phi_{A;B}=
\phi_{\alpha_1\cdots \alpha_\mu;\beta_1\cdots \beta_\nu}\to
e^{2i\bigl(\sum_{\gamma\in A }-\sum_{\gamma\in B }\bigr)
\,\bigl(\sum_j(e_{j\gamma}/m_j)([\cC]_j-[\cB]_j)\bigr)}
\\
\displaystyle
\Psi^*_{\alpha_11}\cdots\Psi^*_{\alpha_\mu1}
\Psi^*_{\beta_12}\cdots\Psi^*_{\beta_\nu2}
\Psi_{\beta_\nu1}\cdots\Psi_{\beta_11}
\Psi_{\alpha_\mu2}\cdots\Psi_{\alpha_12}
\end{array}}
%\label{}
\ee
Then in the 2-point function,
\be
\Braket{0|{\rm T}\,
\phi_{A;B}
(x)\,
\phi^*_{A;B}
(0)|0}
\label{phi2}
\ee
the contribution of the ghost propagators to the power of $x$ at
long distances is
\be
\left(2\sum_{\gamma_1\in A }-2\sum_{\gamma_1\in B }\right)
\left(2\sum_{\gamma_2\in A }-2\sum_{\gamma_2\in B }\right)
\,\frac{1}{2}\lambda^n_{\gamma_1\gamma_2}
\label{gad1}
\ee
\be
=2\left(\sum_{\gamma_1\in A }-\sum_{\gamma_1\in B }\right)
\left(\sum_{\gamma_2\in A }-\sum_{\gamma_2\in B }\right)
\,\left(\delta_{\gamma_1\gamma_2}-\frac{1}{n}u^n_{\gamma_1} u^n_{\gamma_2}\right)
\label{gad2}
\ee
We can now see why overlap between $A$ and
$B$ increases the anomalous dimension.  If there is no
overlap, the first term in the last factor of (\ref{gad2}) gives just
\be
2(\mathbf{N}(A)+\mathbf{N}(B))
\label{nooverlap}
\ee
and
the ghost contribution from this term cancels the contributions from the
free fermion propagators.  Then the scaling at large $x$
comes entirely from the
second term, which because $u^n_\gamma=1$ 
independent of $\gamma$ gives a power
\be
\frac{2\bigl(\mathbf{N}(A)-\mathbf{N}(B)\bigr)^2}{n}
\label{munun}
\ee
The anomalous dimension is half this:
\be
\frac{\bigl(\mathbf{N}(A)-\mathbf{N}(B)\bigr)^2}{n}
\label{ldcad}
\ee
Thus for no overlap, operators with $\mathbf{N}(A)=\mathbf{N}(B)$ are ZDOPs with zero anomalous
dimension.  The rest have anomalous dimensions that go to zero as
$n\to\infty$ for fixed $\mathbf{N}(A)-\mathbf{N}(B)$. 

However, if indices appear in both $A$ and
$B$, 
\be
\mathbf{N}(A\cap B)>0
%\label{}
\ee
then (\ref{nooverlap}) becomes 
\be
2(\mathbf{N}(A)+\mathbf{N}(B))-4\mathbf{N}(A\cap B)
\label{overlap}
\ee
and the 
ghost contributions from the first term 
do not cancel the free-fermion contribution.  Then the
anomalous dimension gets a
positive integer contribution and does not
vanish as $n\to\infty$.
For finite $n$, the operators with $\mathbf{N}(A\cap B)=0$ and
$0<(\mathbf{N}(A)-\mathbf{N}(B))^2<n$ are 
the only scalar 
operators with non-zero anomalous dimensions less than one. 
These are our
LDOPs.\footnote{There is a baryon operator for $n=3$ with dimension $5/6$,
but all the other non-scalar operators have dimension greater than or equal
to one.} 

Thus as $n$ gets large, there are exponentially many
ZDOPs and LDOPs. Their anomalous dimensions depend only
on their chiral charge, $2(\mathbf{N}(A)-\mathbf{N}(B))$, and can be very small even
as the classical dimensions get large!  

\section{Gauge boson masses\label{sec-masses}}

The $\phi$ correlators depend on
dimensional parameters, which must be related to the vector boson masses (or
the related gauge couplings) because these are the only dimensional
parameters in the model.  
The full expression is
\be
\begin{array}{c}
\Braket{0|{\rm T}\,
\phi_{A;B}
(x)\,
\phi^*_{A;B}
(0)|0}
=\biggl(S_1(x)\,
S_2(x)\biggr)^{\mathbf{N}(A)+\mathbf{N}(B)}
\\ \displaystyle
\rule[-5ex]{0ex}{12ex}\times
\prod_{j=1}^{n-1}\exp\left[
\left(\sum_{\gamma_1\in A }-\sum_{\gamma_1\in B }\right)
\left(\sum_{\gamma_2\in A }-\sum_{\gamma_2\in B }\right)
\sum_{j=1}^{n-1}2\lambda_{j\gamma_1\gamma_2}
\left(\kappa(m_j,x)+\ln\left(\sqrt{-x^2 +
i\epsilon}\right)\right)\right]
\end{array}
\label{gadm}
\ee
where
\be
\kappa(m,x)\equiv 
K_0\left(m\sqrt{-x^2 + i\epsilon}\right) + \ln\left(\xi m\right)
\label{kappa}
\ee
Now we can do the sums over $\gamma_1$ and $\gamma_2$
(\ref{ejejgeneral}) and write
\be
\begin{array}{c}
\Braket{0|{\rm T}\,
\phi_{A;B}
(x)\,
\phi^*_{A;B}
(0)|0}
=\biggl(S_1(x)\,
S_2(x)\biggr)^{\mathbf{N}(A)+\mathbf{N}(B)}
\\ \displaystyle
\rule[-5ex]{0ex}{12ex}\times\prod_{j=1}^{n-1}\exp\left[
2h(A,B,j)
\left(
K_0\left(m_j\sqrt{-x^2 + i\epsilon}\right) + \ln\left(\xi m_j\right)
+\ln\left(\sqrt{-x^2 +
i\epsilon}\right)\right)\right]
\end{array}
\label{gadmsums}
\ee
where
{\renewcommand{\arraystretch}{3.5}\be
\begin{array}{c}
\displaystyle
h(A,B,j)\equiv 
\left(\sum_{\gamma_1\in A }-\sum_{\gamma_1\in B }\right)
\left(\sum_{\gamma_2\in A }-\sum_{\gamma_2\in B }\right)
\sum_{j=1}^{n-1}\gamma_{j\gamma_1\gamma_2}
\\ \displaystyle
=\left(N^A_{j+1}+N^B_{j+1}
-\frac{1}{j+1}\left(\sum_{\gamma =1}^{j+1}(N^A_\gamma -N^B_\gamma )\right)^2
+\frac{1}{j}\left(\sum_{\gamma =1}^{j}(N^A_\gamma -N^B_\gamma )\right)^2\right)
\end{array}	
\label{hj}
\ee}
where $N^C_\gamma$ is the number of $\gamma$ indices in the set $C$ (either
0 or 1 because of Fermi statistics): 
\be
N^C_\gamma\equiv \mathbf{N}\bigl(C\cap\{\gamma\}\bigr)
\mbox{~~so that~~
}
\mathbf{N}(C)=\sum_\gamma N^C_\gamma
\label{omega}
\ee

We have already discussed the last term in the exponent in
(\ref{gadmsums}).  It simplifies because
\begin{equation}
\sum_{j-1}^{n-1}h(A,B,j)=
\mathbf{N}(A)+\mathbf{N}(B)-\bigl(\mathbf{N}(A)-\mathbf{N}(B)\bigr)^2/n
\label{hjsum}
\end{equation}
Thus we can write the full result in the form
\be
\left(\frac{\displaystyle\bigl(-x^2 +
i\epsilon\bigr)^{-\bigl(\mathbf{N}(A)-\mathbf{N}(B)\bigr)^2/n}}
{(4\pi^2)^{\mathbf{N}(A)+\mathbf{N}(B)}}\right)
\prod_{j=1}^{n-1}
\left(\xi^2 m_j^2\exp\left[2K_0\left(m_j\sqrt{-x^2+i\epsilon}\right)\right]\right)^{h(A,B,j)}
\label{gadmj}
\ee

For $-x^2\gg1/m_j^2\,\forall\, j$, the $K_0$ factors go to 1 so if we define
\be
H\left(A,B,\{m\}\right)\equiv 
\left(\frac{1}
{(2\pi)^{\mathbf{N}(A)+\mathbf{N}(B)}}\right)
\prod_{j=1}^{n-1}
\left(\xi m_j\right)^{h(A,B,j)}
\label{h}
\ee
then
\be
\Braket{0|{\rm T}\,
\phi_{A;B}
(x)\,
\phi^*_{A;B}
(0)|0}
\mathop{\longrightarrow}_{-x^2\to\infty}
H\left(A,B,\{m\}\right)^2
\left(-x^2+i\epsilon\right)^{-(\mathbf{N}(A)-\mathbf{N}(B))^2/n}
\label{conformalt}
\ee

If $m_j=m\;\forall\, j$, (\ref{gadm}) simplifies to
\be
\begin{array}{c}
\displaystyle
\left(\frac{\xi^2 m^2}{4\pi^2}\right)^{\mathbf{N}(A)+\mathbf{N}(B)}
\left(\xi^2 m^2
(-x^2+i\epsilon)\right)^{-(\mathbf{N}(A)-\mathbf{N}(B))^2/n}
\\ \displaystyle
\times
\exp\left[2\left(\mathbf{N}(A)+\mathbf{N}(B)-(\mathbf{N}(A)-\mathbf{N}(B))^2/n\right)
K_0\left(m\sqrt{-x^2+i\epsilon}\right)\right]
\end{array}
\label{simple}
\ee
so
\be
H\left(A,B,\{m\}\right)\to
\left(\frac{\xi m}{2\pi}\right)^{\mathbf{N}(A)+\mathbf{N}(B)}
\left(\xi m\right)^{-(\mathbf{N}(A)-\mathbf{N}(B))^2/n}
%\label{}
\ee
We will illustrate this before discussing the much
more complicated situation when the masses are different.

Some examples of plots of correlation functions  for equal masses
are shown in figures~\ref{fig-1}
and \ref{fig-2}.
{\figsize\begin{figure}[htb]
$$\includegraphics[width=.6\hsize]{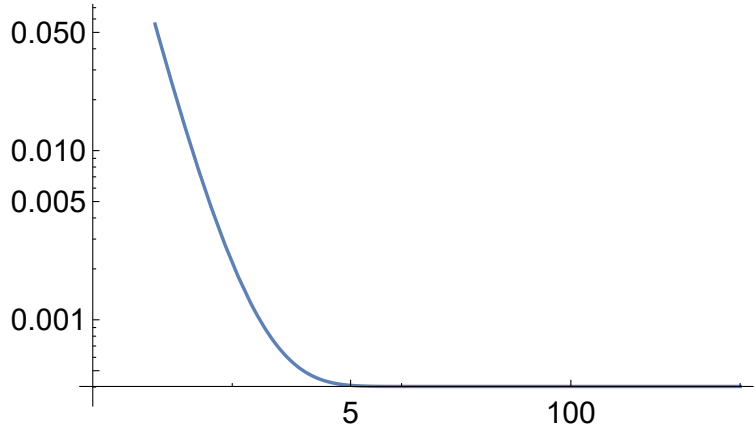}$$
\caption{\figsize\sf\label{fig-1}A ZDOP correlation function for
$\mathbf{N}(A)=\mathbf{N}(B)=1$ as a function of spacelike
$x$ plotted in units of $1/m$, showing the transition from free-fermion
behavior at small $x$ to constancy at large $x$.}\end{figure}}

{\figsize\begin{figure}[htb]
$$\includegraphics[height=.6\hsize]{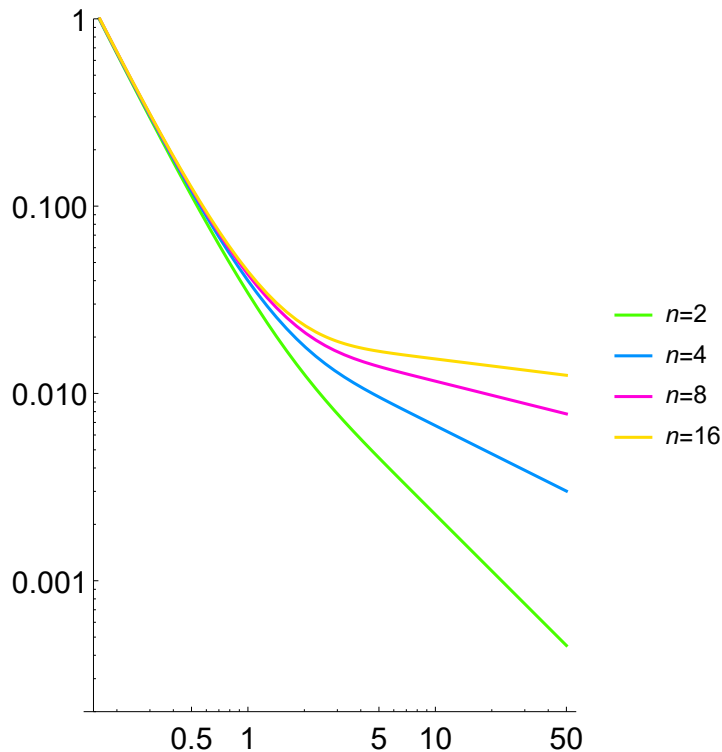}$$
\caption{\figsize\sf\label{fig-2}An LDOP correlation function for
$\mathbf{N}(A)=1$, $\mathbf{N}(B)=0$ as a function of spacelike
$x$ plotted in units of $1/m$ for various values of $n$, showing the
transition from free-fermion 
behavior at small $x$ to unparticle behavior at large $x$.}\end{figure}}

If $\mathbf{N}(A)=\mathbf{N}(B)$, then $\phi_{A;B}$ is a ZDOP 
and we can take $-x^2\to\infty$ in
(\ref{simple}) and use cluster decomposition to conclude that for $m_j=m\,\forall\,j$
\be
\left|\Braket{0|
\phi_{A;B}
|0}\right|^2=H\bigl(A,B,\{m\}\bigr)^2=
\left(\frac{\xi^2 m^2}{4\pi^2}\right)^{\mathbf{N}(A)+\mathbf{N}(B)}
%\label{}
\ee
so that 
\be
\Braket{0|
\phi_{A;B}
|0}=H\bigl(A,B,\{m\}\bigr)\,e^{i\theta_{A;B}}
\label{zdopvev}
\ee
Thus, just as in the Schwinger model, the vacuum is degenerate.  Here we
would expect that there will be $n$$-$$1$ independent $\theta$ angles, one
for each gauge boson and each (related) anomalous $U(1)$ symmetry.  We will
get further insight into this in section~\ref{sec-3pt}.

The relations we have derived for equal couplings and masses 
allow us to discuss some of the physics of these models simply.  But we are also
interested in thinking about what happens when there
are multiple gauge boson masses, and in particular, whether there is
anything interesting about how the long-distance limit is approached. 

We will illustrate the general result (\ref{gadmj}) with three masses.
With different masses,
we can no longer shuffle the gauge couplings willy-nilly.  But we can give a
sense of some of the structure of (\ref{gadmj}) by discussing 
figures~\ref{fig-3}-\ref{fig-6}.  At distances small compared
to all the gauge boson Compton wavelengths, we see the free fermion
behavior, depending only on 
the total number of fermion fields in the operator.  But as
the separation grows, the correlators split into branches depending on the
parameters $m_j$, $N^A_\gamma$ and $N^B_\gamma$ . 
In figures~\ref{fig-3}-\ref{fig-6}, we look at $n=4$ diagonal color with
three different masses, $m$, $10m$, and $100m$ and plot the correlators for
various $\phi_{A;B}$ versus $x$ in units of $1/m$. The legends show
\be
m\mu\nu(\{m_1/m,m_2/m,m_3/m\},
\{N^A_1,N^A_2,N^A_3,N^A_4\},\{N^B_1,N^B_2,N^B_3,N^B_4\})
\label{legends}
\ee
to make it easier see the dependence on the order of the $m$s.

Figure~\ref{fig-3} shows the correlator for an LDOP with $N^B_3=1$ and other
$\mu$s and $\nu$s zero in the notation of (\ref{gadmj}).\footnote{Note
that the aspect ratio changes in figure~\ref{fig-3}-\ref{fig-6} }  You can see
the free fermion behavior at low $x$ splitting into various different
branches and becoming parallel at large $x$ where there the approximate
conformal symmetry takes over and the scaling behavior is fixed by the
anomalous dimension.  Because $N^B_3=1$ in figure~\ref{fig-3} with other
$\mu$s and $\nu$s zero, the largest effect 
comes from the second term in the $j=2$ term in  (\ref{gadmj}) and
occurs at
$x\approx1/m_2$ where the $\cC_2$ ghost field begins to cancel the free
fermion scaling.  Thus the top 2 graphs with $m_2=100m$ diverge from
free-fermion behavior first at
$x\approx.01$, the 
middle 2 with $m_2=10m$ diverge next at $x\approx.1$, and finally the last
two with $m_2=m$ diverge at $x\approx1$.  The smaller spittings come
from the anomalous dimension terms (the $1/(j+1)$ and $1/j$ terms in
(\ref{gadmj})).
{\figsize\begin{figure}[htb]
$$\includegraphics[width=.99\hsize]{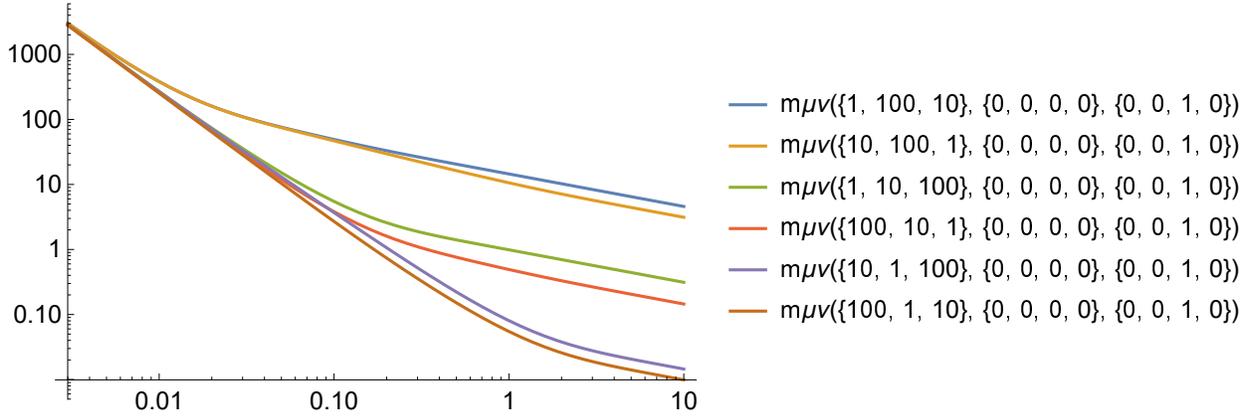}$$
\caption{\figsize\sf\label{fig-3}LDOP correlators for various permutations
of $(1,10,100)\to\left(\frac{m_1}{m},\frac{m_2}{m},\frac{m_3}{m}\right)$.}\end{figure}}

Similar patterns appear in the ZDOP correlators in
figures~\ref{fig-4}-\ref{fig-6}, but they are more 
complicated because more masses are involved in the largest effects. 

{\figsize\begin{figure}[htb]
$$\includegraphics[width=.99\hsize]{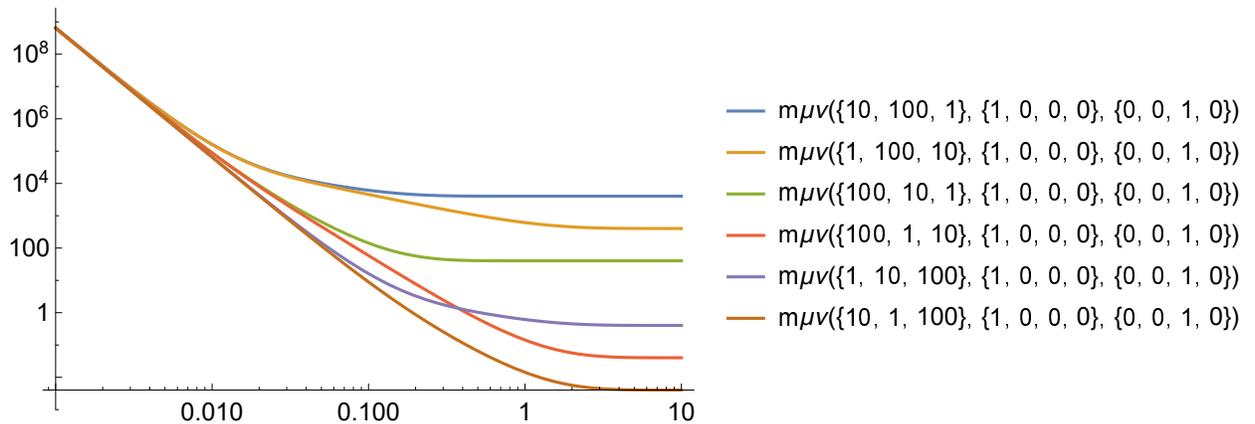}$$
\caption{\figsize\sf\label{fig-4}ZDOP correlators for various permutations
of $(1,10,100)\to\left(\frac{m_1}{m},\frac{m_2}{m},\frac{m_3}{m}\right)$.}\end{figure}}

{\figsize\begin{figure}[htb]
$$\includegraphics[width=.99\hsize]{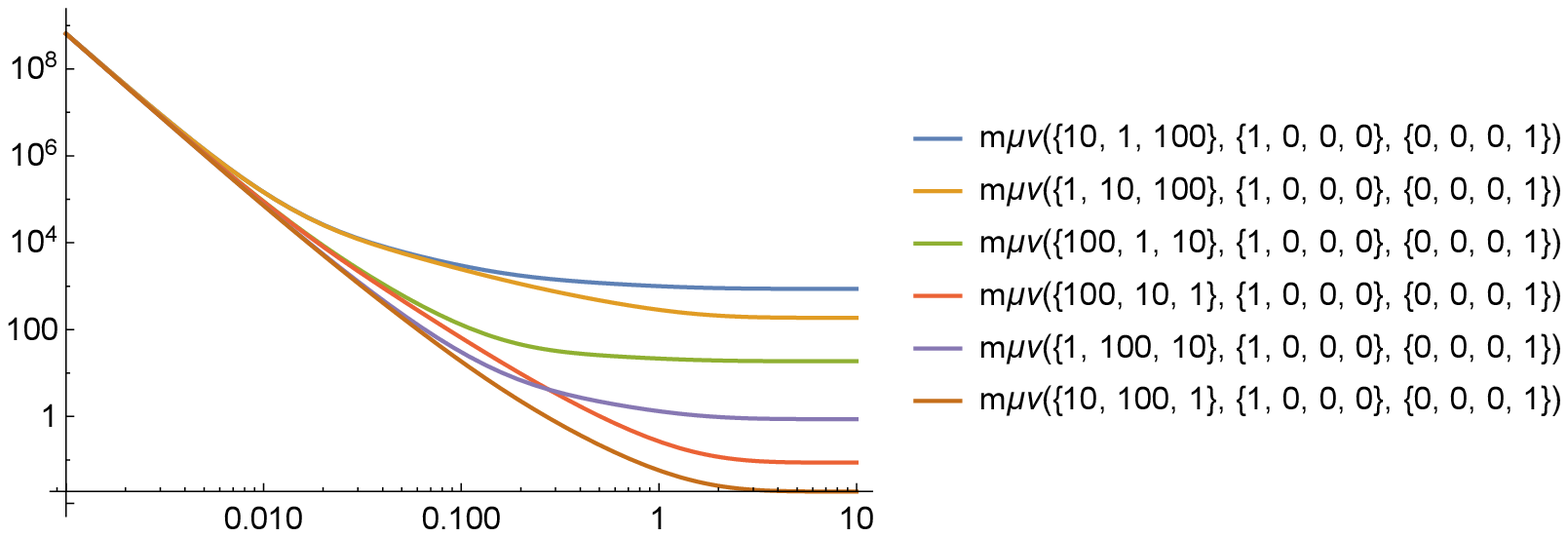}$$
\caption{\figsize\sf\label{fig-5}ZDOP correlators for various permutations
of $(1,10,100)\to\left(\frac{m_1}{m},\frac{m_2}{m},\frac{m_3}{m}\right)$.}\end{figure}}

{\figsize\begin{figure}[htb]
$$\includegraphics[width=.99\hsize]{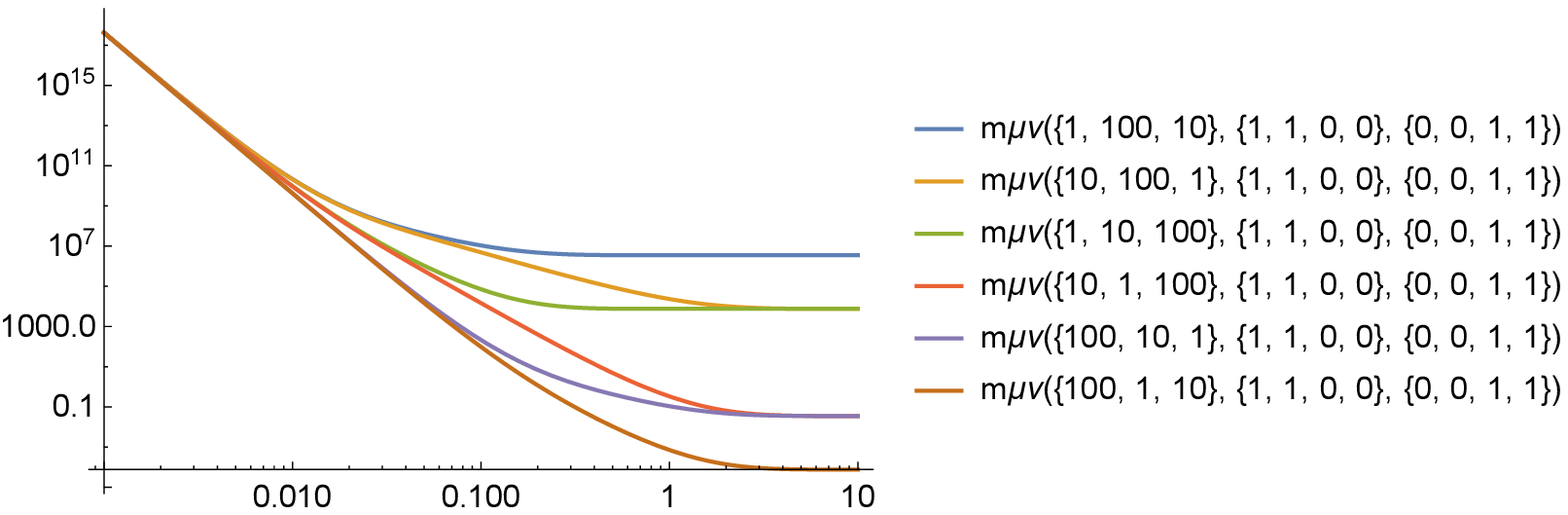}$$
\caption{\figsize\sf\label{fig-6}ZDOP correlators for various permutations
of $(1,10,100)\to\left(\frac{m_1}{m},\frac{m_2}{m},\frac{m_3}{m}\right)$.}\end{figure}}

You can see from figures~\ref{fig-4} and \ref{fig-5} that the ZDOP VEVs
depend not just on the gauge boson masses but on where they appear in $A$
and $B$, as expected from (\ref{hj}) and (\ref{h}).
For $\mathbf{N}(A)=1$, it is manageable (though not particularly
edifying) to write down the VEVs explicitly.  This is done in
appendix~\ref{sec-vevmasses}. 

We should also emphasize that the results of this section depend on the
details of the diagonal color couplings, (\ref{e-diagonal-color}).  As we
discussed in section~\ref{sec-zdops}, we can construct a generalized
Schwinger model for any set of diagonal couplings, $e_{j\alpha}$ of the
gauge bosons $A^\mu_j$ to the fermion fields $\psi_\alpha$ satisfying
(\ref{m02}), $m_j\delta_{jk}=\sum_\alpha e_{j\alpha}e_{k\alpha}/\pi$.  For
any set of $n$ fermions and $n$$-$$1$ gauge bosons satisfying (\ref{m02}),
we can write
\begin{equation}
\sum_{j=1}^{n-1}\frac{e_{j\alpha}e_{j\beta}}{\pi m_j^2}
=\delta_{\alpha\beta}-v_\alpha v_\beta
\label{lambdageneral}
\end{equation}
for some unit vector $v_\alpha$.  Just as for diagonal color, when we compute
the 2-point function (\ref{phi2}) for $\phi_{A;B}$ satisfying
(\ref{acapb}), the $\delta$-function term in 
(\ref{lambdageneral}) gives a ghost contribution
that cancels the free-fermion contribution to the $x$ dependence, but
generically, the $v_\alpha v_\beta$ term will give a non-zero contribution
to the anomalous dimension.  Thus generically there will be no ZDOPs.  

Diagonal color is special because of (\ref{eoverm}) which fixes the
vector $v_\alpha$ to be
\be
v_\alpha=\frac{1}{\sqrt{n}}u^n_\alpha
%\label{}
\ee
But in fact, this is not that special!  Any set of $e_{j\alpha}$ satisfying
(\ref{m02}) and orthogonal to $u^n_\alpha$ 
\be
\sum_\alpha e_{j\alpha}u^n_\alpha
=\sum_\alpha e_{j\alpha}
=0
\label{baryons}
\ee
will satisfy (\ref{eoverm}), and
will have the same ZDOPs, LDOPs, and LDOP anomalous dimensions as diagonal
color.  But the dependence on the gauge boson masses will be completely different.
Notice that (\ref{baryons}) is the condition that the model contains
gauge-invariant baryonic unparticle operators with one $\psi_{1\alpha}$ or 
$\psi_{2\alpha}$ for each $\alpha$. 

It is also worth noting that among the infinite possibilities for couplings
satisfying (\ref{m02}) and (\ref{baryons}) is one that is well-beloved by all
physicists - the oscillatory normal modes of identical masses connected by
identical springs on a circle (or on a line with periodic boundary conditions)
are orthogonal to the translational mode - which is just the condition
(\ref{baryons}).  For example, for $n$ odd we could have
\be
e_{j\alpha}=\sqrt{\frac{2\pi}{n}}m_j\cos\bigl(2\pi j\alpha/n\bigr)
\quad
e_{(j+\frac{n-1}{2})\alpha}=
\sqrt{\frac{2\pi}{n}}m_{j+\frac{n-1}{2}}\sin\bigl(2\pi j\alpha/n\bigr)
\mbox{~~for $j=1$ to $(n-1)/2$.}
%\label{}
\ee
A model with these couplings would be very different from
diagonal color for different gauge boson masses because the modes
are evenly spread over all the fermions.

\section{Perturbative 3-point functions\label{sec-3pt}}

In the notation of (\ref{phi}),
We now consider the 3-pt function
\be
\Braket{0|{\rm T}\,
\phi_{A_1;B_1}
(x_1)\,
\phi_{A_2;B_2}
(x_2)\,
\phi_{A_3;B_3}
(x_3)
|0}
\label{phi3}
\ee
where the $\phi$s are ZDOPs or LDOPS (which therefore satisfy (\ref{acapb})).
The correlator (\ref{phi3}) satisfying (\ref{acapb}) and (\ref{klnooverlap})
is non-zero in
perturbation theory if and only if  all the fermion-antifermion pairs can
combine into propagators. This requires
\be
\sum_kN^{A_k}_\gamma
=
\sum_kN^{B_k}_\gamma\;\;\forall\,\gamma
\label{combine}
\ee
For three operators, the conditions (\ref{acapb}) and (\ref{combine}) together with the
exclusion principle implies
\be
A_k\cap A_\ell=B_k\cap B_\ell=\varnothing\mbox{~for $k\neq\ell$.}\\
\label{klnooverlap}
\ee
because if two of the $A$s
or $B$s contained an index twice, 
the third $B$ or $A$ 
would have to contain the index twice, and the operator would
vanish.  Thus we conclude that\footnote{See also (\ref{uaub}) below.} 
\be
A_1\cup A_2\cup A_3=B_1\cup B_2\cup B_3
\label{sets}
\ee 
Then the correlator is
\be
\begin{array}{c}
\displaystyle
\Braket{0|{\rm T}\,
\phi_{A_1;B_1}
(x_1)\,
\phi_{A_2;B_2}
(x_2)\,
\phi_{A_3;B_3}
(x_3)
|0}
=\prod_{k<\ell}
\left\{\rule{0ex}{5ex}\biggl(S_1(x_k-x_\ell)\,
S_2(x_k-x_\ell)\biggr)^{N^{AB}_{k\ell}+N^{AB}_{\ell k}}\right.
\\ \displaystyle
\left.\times\prod_{j=1}^{n-1}\exp\left[
2h(A_k,B_k,A_\ell,B_\ell,j)
\left(
\kappa(m_j,x_k-x_\ell)
+\ln\left(\sqrt{-(x_k-x_\ell)^2 +
i\epsilon}\right)\right)\right]\rule{0ex}{5ex}\right\}
\end{array}
\label{gad3}
\ee
where\footnote{In section~\ref{sec-rpoint} we will need $N^{AA}$ and $N^{BB}$ so
we keep the sets 
arbitrary in (\ref{mukell}).}
\be
N^{CD}_{k\ell}\equiv \mathbf{N}\bigl(C_k\cap D_\ell)
\label{mukell}
\ee
and
{\renewcommand{\arraystretch}{3}
\be
\begin{array}{c}
\displaystyle
h(A_k,B_k,A_\ell,B_\ell,j)\equiv 
-
\left(\sum_{\gamma_1\in A_k}-\sum_{\gamma_1\in B_k}\right)
\left(\sum_{\gamma_2\in A_\ell}-\sum_{\gamma_2\in B_\ell}\right)
\lambda_{j\gamma_1\gamma_2}
\\ \displaystyle
=
-\left(N^{A_k}_{j+1}-N^{B_k}_{j+1}\right)\left(N^{A_\ell}_{j+1}-N^{B_\ell}_{j+1}\right)
+\frac{1}{j+1}
\left(\sum_{\gamma_1 =1}^{j+1}(N^{A_k}_{\gamma_1} -N^{B_k}_{\gamma_1} )\right)
\left(\sum_{\gamma_2 =1}^{j+1}(N^{A_\ell}_{\gamma_2} 
-N^{B_\ell}_{\gamma_2} )\right)
\\ \displaystyle
-\frac{1}{j}
\left(\sum_{\gamma_1 =1}^{j}(N^{A_k}_{\gamma_1} -N^{B_k}_{\gamma_1} )\right)
\left(\sum_{\gamma_2 =1}^{j}(N^{A_\ell}_{\gamma_2} 
-N^{B_\ell}_{\gamma_2} )\right)
\end{array}	
\label{hkell}
\ee}%
The minus sign relative to (\ref{hj}) arises because none
of these $\phi$s are conjugated.  In this application, 
\be
N^{A_k}_{j+1}N^{A_\ell}_{j+1}=N^{B_k}_{j+1}N^{B_\ell}_{j+1}=0
%\label{}
\ee
because of (\ref{klnooverlap}), but (\ref{hkell}) is entirely general.
Note that (see (\ref{acapb}))
\be
h(A,B,j)=h(A,B,B,A,j)=-h(A,B,A,B,j)
\label{habhabba}
\ee

As usual, the long distance $x$ dependence is easy to deal with because we
can do the sum over $j$ and use (\ref{kappa}) to write (\ref{gad3}) as  
\be
\begin{array}{c}
\displaystyle
\prod_{k<\ell}
\left\{\rule{0ex}{5ex}
\left(\frac{1}{4\pi^2}\right)^{N^{AB}_{k\ell}+N^{AB}_{\ell k}}
\left(-(x_k-x_\ell)^2 +
i\epsilon\right)^{
\bigl(\mathbf{N}(A_k)-\mathbf{N}(B_k)\bigr)
\bigl(\mathbf{N}(A_\ell)-\mathbf{N}(B_\ell)\bigr)/n
}\right.
\\ \displaystyle
\left.\times\prod_{j=1}^{n-1}
\left(\xi^2m_j^2\exp\left[
2
K_0\left(m_j\sqrt{-(x_k-x_\ell)^2+i\epsilon}\right)
\right]\right)^{h(A_k,B_k,A_\ell,B_\ell,j)}
\rule{0ex}{5ex}\right\}
\end{array}
\label{gad3x}
\ee
Notice that because of (\ref{combine}) and (\ref{omega}) we know that
\be
\sum_k\Bigl(\mathbf{N}(A_k)-\mathbf{N}(B_k)\Bigr)=0
%\label{}
\ee
and therefore the total scaling dimension of (\ref{gad3x})
\be
2\sum_{k<\ell}\Bigl(\mathbf{N}(A_k)-\mathbf{N}(B_k)\Bigr)
\Bigl(\mathbf{N}(A_\ell)-\mathbf{N}(B_\ell)\Bigr)/n
=-\sum_k\Bigl(\mathbf{N}(A_k)-\mathbf{N}(B_k)\Bigr)^2/n
%\label{}
\ee
which is just the sum of the anomalous dimensions of the $\phi$s.

Finally, notice that the correlator is positive for space-like separations.
This sign comes from keeping track
of Fermi statistics of the fermions and we get a positive sign because of
our definition of the order of the indices in $\phi$ in (\ref{phi}) (see
appendix~\ref{sec-signs}).   

The form (\ref{gadmj}) for the perturbative 2-point functions 
implies that if we define the normalized operators
\be
\Phi_{A;B}\equiv 
(4\pi^2)^{\mathbf{N}(A)+\mathbf{N}(B)}\phi_{A;B}/H\bigl(A,B,\{m\}\bigr)
\label{Phi}
\ee
the long-distance part has a simple form
\be
%\begin{array}{c}
\Braket{0|{\rm T}\,
\Phi_{A;B}
(x)\,
\Phi^*_{A;B}
(0)\,
|0}
%\\
\displaystyle
\mathop{\longrightarrow}_{-x^2\to\infty}
\,\left(
-x^2+i\epsilon\right)^{-(\mathbf{N}(A_1)-\mathbf{N}(B_1))^2/n}
%\end{array}
\label{flow2}
\ee
The form (\ref{gad3x}) then implies that the 3-point functions of the
normalized operators are equally simple.
To see this, note that to have a non-zero 3-point function in perturbation
theory we must have (\ref{combine}) which
implies
\be
\sum_{\ell}
\left(\sum_{\gamma_1\in A_k}-\sum_{\gamma_1\in B_k}\right)
\left(\sum_{\gamma_2\in A_\ell}-\sum_{\gamma_2\in B_\ell}\right)
\lambda_{j\gamma_1\gamma_2}
=0
\label{setszero}
\ee
because the sums over $\gamma_2$ in the $A$s cancel the sums over
$\gamma_2$ in the $B$s.  This implies that 
\be
h(A_k,B_k,A_k,B_k,j)+\sum_{\ell\neq k}h(A_k,B_k,A_\ell,B_\ell,j)=0
\label{hkhk}
\ee
and thus using (\ref{habhabba}) and symmetry of
$h(A_k,B_k,A_\ell,B_\ell,j)$ under $k\leftrightarrow\ell$,
\be
\sum_{k<\ell}h(A_k,B_k,A_\ell,B_\ell,j)=\frac{1}{2}\sum_{k}h(A_k,B_k,j)
\label{hkhkell}
\ee
This implies that when all the distances go to spacelike 
infinity\footnote{The factors of $4\pi^2$ always work out properly just
by counting because they
come from the free-fermion skeleton.}
\be
\begin{array}{c}
\Braket{0|{\rm T}\,
\Phi_{A_1;B_1}
(x_1)\,
\Phi_{A_2;B_2}
(x_2)\,
\Phi_{A_3;B_3}
(x_3)\,
|0}
\\
\displaystyle
\mathop{\longrightarrow}
\sum_{k<\ell}\left(
-(x_k-x_\ell)^2+i\epsilon\right)^{\bigl(\mathbf{N}(A_k)-\mathbf{N}(B_k)\bigr)
\bigr(\mathbf{N}(A_\ell)-\mathbf{N}(B_\ell)\bigr)/n}
\end{array}
\label{flow3}
\ee

We will see in the next two sections that (\ref{flow2}) and (\ref{flow3})
generalize to arbitrary perturbative {\bf and non-perturbative} correlators.

For a ZDOP, (\ref{flow2}) implies that its VEV is 1. 
If the three operators in the 3-pt function (\ref{flow3}) are ZDOPs, then
taking the distances to spacelike infinity we conclude
\be
\Braket{0|
\Phi_{A_1;B_1}
|0}
\Braket{0|
\Phi_{A_2;B_2}
|0}
\Braket{0|
\Phi_{A_3;B_3}
|0}=1
\label{rgad3kellxi}
\ee
The magnitude of (\ref{rgad3kellxi}) does not tell us anything we did not
already know from (\ref{flow2}), but the sign has
consequences.
Comparing with (\ref{zdopvev}), (\ref{rgad3kellxi}) shows that for 
$A_j$ and $B_j$ satisfying 
(\ref{acapb}),
(\ref{klnooverlap}), and  (\ref{sets}), the phases satisfy
\be
e^{i\theta_{A_1;B_1}}
e^{i\theta_{A_2;B_2}}
e^{i\theta_{A_3;B_3}}=1
\label{phaserel}
\ee
We also know from (\ref{phistar}) that
\be
e^{-i\theta_{A;B}}=e^{i\theta_{B;A}}
%\label{}
\ee
Thus while the phase of any particular ZDOP VEV is arbitrary because of vacuum
degeneracy, there are relations among the phases of different VEVs.
Using (\ref{phaserel}) with (\ref{klnooverlap}) and (\ref{sets}), we can
determine the phases of 
all the ZDOP VEVs in diagonal color $SU(n)$ in terms of $n$$-$1 independent
phases, which we can take 
to be
\be
\theta_j\equiv \theta_{\{j\};\{n\}}\mbox{~for $j=1$ to $n$$-$1}
\label{thetaj}
\ee
Then  (\ref{phaserel}) implies
\be
\theta_{A;B}=\sum_{\alpha\in A}\theta_\alpha-\sum_{\alpha\in B}\theta_\alpha
=\sum_\alpha\bigl(N^A_\alpha-N^B_\alpha\bigr)\theta_\alpha
\mbox{~where $\theta_n\equiv 0$.}
\label{thetaab}
\ee
Having one independent $\theta$ angle for each of the $n$$-$1 gauge fields
is consistent with the connection between $\theta$ angles and
background electric fields~\cite{Coleman:1975pw,Coleman:1976uz}. 

The form (\ref{thetaab}) also means that we can always redefine our $\Phi$s so
that the VEV's are real.  This is not surprising because the $\theta$ vacua are
completely degenerate for massless fermions.  We will assume that we have
done this and that for all ZDOPs
\be
\Braket{0|
\Phi_{A;B}
|0}
=1
\label{Phiz}
\ee

\section{Perturbative $\mathbf{r}$-point functions\label{sec-rpoint}}

Now we will find all the perturbative correlators of ZDOPs and LDOPs.  We
will then be able to use the perturbative calculation along with cluster
decomposition to find all the correlators, perturbative and non-perturbative.
We have already done most of the work in section~\ref{sec-3pt}.
Again we use
the notation of (\ref{phi}) and write our ZDOPs and LDOPs as $\phi_{A;B}$
where $A$ and $B$ are sets of fermion color indices where
\be
A_k\cap B_k=\varnothing\,\forall\, k
\label{rsetsnooverlap}
\ee
The correlator
\be
\Braket{0|{\rm T}\,
\prod_{k=1}^r\phi_{A_k;B_k}
(x_k)
|0}
\label{phir}
\ee
is nonzero in
perturbation theory if and only if  all the fermion-antifermion pairs with
every index can
combine into propagators).
Thus we require\footnote{Recall that for $r=3$, because each index appears at
most once in the $A$s and in the $B$s, this condition was equivalent to
(\ref{sets}), but (\ref{uaub}) is the general form.  }
\be
\sum_{k=1}^rN^{A_k}_\gamma
=
\sum_{k=1}^rN^{B_k}_\gamma
\;\forall\,\gamma
\label{uaub}
\ee
For $r>3$, the exclusion principle does {\bf not} require 
\be
A_k\cap A_\ell=B_k\cap B_\ell=\varnothing\mbox{~for $k\neq\ell$.}\\
\label{rklnooverlap}
\ee
so there are additional contributions to the fermion skeleton that make use
of the magic of 1D fermions 
\be
\left(\sum_{P}(-1)^{s(P)}\prod_{j=1}^{n} S_\ell (x_{j}-y_{P(j)})\right)
=(-1)^{n(n-1)/2}\prod_{j,k=1}^{n} S_\ell (x_{j}-y_{k})/\prod_{j<k} S_\ell
(x_{j}-x_{k})/\prod_{j<k} S_\ell (y_{j}-y_{k})
\label{magic}
\ee
and the correlator is
\be
\Braket{0|{\rm T}\,
\prod_{k=1}^r\phi_{A_k;B_k}
(x_k)
|0}
=
\prod_{k<\ell}\cD_{k\ell}
\label{rgad3}
\ee
where\footnote{This looks like (\ref{gad3}) except that in (\ref{gad3})
there are no $N^{AA}$ and $N^{BB}$ terms because of the
condition (\ref{klnooverlap}) which follows from (\ref{acapb}) and
the exclusion principle.}
\be
\begin{array}{c}
\cD_{k\ell}\equiv \displaystyle\biggl(S_1(x_k-x_\ell)\,
S_2(x_k-x_\ell)\biggr)^{N^{AB}_{k\ell}+N^{AB}_{\ell k}-N^{AA}_{k\ell}-N^{BB}_{k\ell}}
\\ \displaystyle
\left.\times\prod_{j=1}^{n-1}\exp\left[
2h(A_k,B_k,A_\ell,B_\ell,j)
\left(
\kappa(m_j,x_k-x_\ell)
+\ln\left(\sqrt{-(x_k-x_\ell)^2 +
i\epsilon}\right)\right)\right]\rule{0ex}{5ex}\right\}
\end{array}
\label{rgad3kell}
\ee
Again we can simplify the $x$ dependence and write this as
\be
\begin{array}{c}
\displaystyle
\cD_{k\ell}=\prod_{k<\ell}
\left\{\rule{0ex}{5ex}
\left(\frac{1}{4\pi^2}\right)^{N^{AB}_{k\ell}+N^{AB}_{\ell k}
-N^{AA}_{k\ell}-N^{BB}_{k\ell}}
\left(-(x_k-x_\ell)^2 +
i\epsilon\right)^{
\bigl(\mathbf{N}(A_k)-\mathbf{N}(B_k)\bigr)
\bigl(\mathbf{N}(A_\ell)-\mathbf{N}(B_\ell)\bigr)/n
}\right.
\\ \displaystyle
\left.\times\prod_{j=1}^{n-1}
\bigl(\xi^2m_j^2\bigr)^{h(A_k,B_k,A_\ell,B_\ell,j)}
\exp\left[
2h(A_k,B_k,A_\ell,B_\ell,j)
K_0\left(m_j\sqrt{-(x_k-x_\ell)^2+i\epsilon}\right)
\right]\rule{0ex}{5ex}\right\}
\end{array}
\label{rgad3x}
\ee

The same argument that gives (\ref{hkhk}) and (\ref{hkhkell}) implies that
the normalized $\Phi$s have simple correlators
\be
%\begin{array}{c}
\Braket{0|{\rm T}\,
\prod_{k=1}^r\Phi_{A_k;B_k}
(x_k)
|0}
%\\
\displaystyle
\mathop{\longrightarrow}_{-x^2\to\infty}
\prod_{k<\ell}\left(
-(x_k-x_\ell)^2+i\epsilon\right)^{\bigl(\mathbf{N}(A_k)-\mathbf{N}(B_k)\bigr)
\bigr(\mathbf{N}(A_\ell)-\mathbf{N}(B_\ell)\bigr)/n}
%\end{array}
\label{flowr}
\ee

\section{Non-perturbative 2-point functions and
coalescence\label{sec-coalescence}} 

Now we work with the normalized operators defined by (\ref{Phi}) and 
and (\ref{Phiz})
restrict ourselves to a pair $\Phi_{A_1,B_1}$  and $\Phi_{A_2,B_2}$ 
with
\be
\mathbf{N}(A_1)
-\mathbf{N}(B_1)
=\mathbf{N}(A_2)
-\mathbf{N}(B_2)
%\label{}
\ee
so the two $\Phi$s have the same chiral charges and (therefore) the same
anomalous dimension.
Because the 2-point function 
\be
\Braket{0|{\rm T}\,
\Phi_{A_1;B_1}
(x)\,
\Phi^*_{A_2;B_2}
(0)\,
|0}
\label{2ptgeneral}
\ee
is not forbidden by any symmetry that survives
in the quantum theory and because anything that can happen usually does, we
expect that it will be non-zero, even if $\{A_1;B_1\}\neq\{A_2;B_2\}$ so
that it vanishes in perturbation theory.
We can calculate the absolute value of (\ref{2ptgeneral})
by applying cluster decomposition to the perturbatively calculable 4-point function
\be
\begin{array}{c}
\displaystyle\Braket{0|{\rm T}\,
\Phi_{A_1;B_1}
(x_1)\,
\Phi^*_{A_1;B_1}
(x_2)\,
\Phi_{A_2;B_2}
(x_3)\,
\Phi^*_{A_2;B_2}
(x_4)\,
|0}
\\
\displaystyle
=\Braket{0|{\rm T}\,
\Phi_{A_1;B_1}
(x_1)\,
\Phi_{B_1;A_1}
(x_2)\,
\Phi_{A_2;B_2}
(x_3)\,
\Phi_{B_2;A_2}
(x_4)\,
|0}
\end{array}
\label{4point12}
\ee
We can calculate this exactly using the results of section~\ref{sec-rpoint}
but what we are most interested in is the behavior of (\ref{4point12})
at long distances, and for that
the only thing that we will need from the result is the $x$ dependence.  It
follows from (\ref{rgad3kell}) and (\ref{rgad3kellxi}) that
(\ref{4point12}) is
\be
\left(\frac{\displaystyle\prod_{k-\ell\;\rm even}\Bigl(-(x_k-x_\ell)^2+i\epsilon\Bigr)}
{\displaystyle\prod_{k-\ell\;\rm
odd}\Bigl(-(x_k-x_\ell)^2+i\epsilon\Bigr)}\right)^{(\mathbf{N}(A_1)
-\mathbf{N}(B_1))^2/n}
\label{c4point12}
\ee
times exponentials that go to $1$ at large distances.
Now we can apply cluster decomposition
\be
\begin{array}{c}
\displaystyle
\Braket{0|{\rm T}\,
\Phi_{A_1;B_1}
(x_1)\,
\Phi^*_{A_1;B_1}
(x_2+z)\,
\Phi_{A_2;B_2}
(x_3+z)\,
\Phi^*_{A_2;B_2}
(x_4)\,
|0}
\\
\displaystyle
\mathop{\longrightarrow}_{-z^2\to\infty}
\Braket{0|{\rm T}\,
\Phi_{A_1;B_1}
(x_1)\,
\Phi^*_{A_2;B_2}
(x_4)\,
|0}
\Braket{0|{\rm T}\,
\Phi_{A_2;B_2}
(x_3)\,
\Phi^*_{A_1;B_1}
(x_2)\,
|0}
\\
\displaystyle
=\biggl(\Bigl(-(x_1-x_4)^2+i\epsilon\Bigr)
\Bigl(-(x_3-x_2)^2+i\epsilon\Bigr)\biggr)^{-(\mathbf{N}(A_1)
-\mathbf{N}(B_1))^2/n}
\end{array}
\label{z23}
\ee
(\ref{z23}) implies that up to a phase
\be
\Braket{0|{\rm T}\,
\Phi_{A_1;B_1}
(x)\,
\Phi^*_{A_2;B_2}
(0)\,
|0}
=\bigl(-x^2+i\epsilon\bigr)^{-(\mathbf{N}(A_1)
-\mathbf{N}(B_1))^2/n}
\label{z23i}
\ee

We can determine the phase as well because we can always
find ZDOPs $\Phi_{A_k;B_k}(x_k)$ such that
\be
\Braket{0|{\rm T}\,
\Phi_{A_1;B_1}
(x)\,
\Phi_{B_2;A_2}
(0)\,\prod_{k=3}^s\Phi_{A_k;B_k}(x_k)
|0}
\label{np2zdops}
\ee
is non-zero in pertubation theory. 
If there are no repeat indices, then a single $\Phi_{B_1\cup B_2;A_1\cup
A_2}$ does the job.  If there are repeat indices, then we can break this up
into subsets satisfying the ZDOP constraints (it is sometimes necessary to
add additional pairs of indices).
Then we can get the phase by combining
the sign of the perturbative correlator from (\ref{rgad3}) with the ZDOP
phases from (\ref{thetaab}).  In practice, we do not need to actually find examples 
of the $\Phi_{A_k;B_k}(x_k)$ in (\ref{np2zdops}) 
because all such correlators give the same result. There are no phases!

Thus we conclude that 
the long-distance correlators 
of normalized LDOPs depend only their chiral charge.
\be
%\begin{array}{c}
\Braket{0|{\rm T}\,
\Phi_{A_1;B_1}
(x)\,
\Phi^*_{A_2;B_2}
(0)\,
|0}
%\\
\displaystyle
\mathop{\longrightarrow}_{-x^2\to\infty}\delta_{\mathbf{N}(A_1)- \mathbf{N}(B_1),
\mathbf{N}(A_2)- \mathbf{N}(B_2)}
\,\left(
-x^2+i\epsilon\right)^{-(\mathbf{N}(A_1)-\mathbf{N}(B_1))^2/n}
%\end{array}
\label{flow2np}
\ee

It is straightforward to extend these arguments to arbitrary correlators
with the result
{\renewcommand{\arraystretch}{3}\be
\begin{array}{c}
\displaystyle
\Braket{0|{\rm T}\,
\prod_{k=1}^r\Phi_{A_k;B_k}
(x_k)\,
|0}
\mathop{\longrightarrow}_{-x^2\to\infty}
\delta_{\left(\sum_k\mathbf{N}(A_k)\right),\left(\sum_k\mathbf{N}(B_k) \right)}
\\ \displaystyle
\times\prod_{k<\ell}\left(-(x_k-x_\ell)^2+i\epsilon\right)^{\bigl(\mathbf{N}(A_k)
-\mathbf{N}(B_k)\bigr)\bigl(\mathbf{N}(A_\ell)-\mathbf{N}(B_\ell)\bigr)/n}
\end{array}
\label{flowr2}
\ee}% 
This is an extreme form of conformal coalescence.  All differences of $\Phi$s
with the same chiral charge vanish exponentially for $x\gg 1/m$.  
(\ref{flow2np}) and (\ref{flowr2}) are consistent with the hypothesis that all the
$\Phi$s with the same chiral charge
flow to the same operator in the conformal sector.  
\be
\Phi_{A;B}(x)
\mathop{\longrightarrow}_{-x^2\to\infty}
O_{\mathbf{N}(A)- \mathbf{N}(B)}(x)
\label{flow}
\ee
One might say that
there is only one kind of unparticle stuff for each chiral charge.  However,
the details of how the limit is approached depend on the operator and the
boson masses.

The result (\ref{flowr2}) does not depend on all the details of diagonal
color.  The results of sections \ref{sec-3pt}-\ref{sec-coalescence} are
valid for any set of gauge couplings satisfying (\ref{eoverm}) and any gauge
boson masses.  The precise form of the normalization factors will depend on
the details, but coalescence is a very general result.

\section{Incomplete binding\label{sec-ib}}

We want to close with some speculations about what is going on here.  We
suspect that there
is a sense in which unparticle behavior, at least in these 1+1 dimensional
models, represents the theory's failed attempt to completely bind the massless
fermions into gauge-invariant particle states.  The unparticle LDOPs are the next
best thing, intermediate between massive particles and collections of
massless particles!~\cite{Georgi:2007ek}  The ZDOPs chronicle the theory's
progress towards complete binding.  One way of attempting to quantify
this suspicion is by adding 
a $U(1)$ gauge coupling with a very small scale, $m_n$.  Then (\ref{ejejgeneral}) and
(\ref{gadmj}) become 
\be
\frac{e_{j{\gamma_1}}e_{j{\gamma_2}}}{\pi m_j}
=
\left\{
\begin{array}{l}
\displaystyle
\delta_{{\gamma_1},j+1}\delta_{{\gamma_2},j+1}
-\frac{1}{j+1}u^{j+1}_{\gamma_1} u^{j+1}_{\gamma_2}
+\frac{1}{j}u^{j}_{\gamma_1} u^{j}_{\gamma_2}\mbox{~~for $j=1$ to $n$$-$$1$}
\\
\displaystyle
\frac{1}{n}u^{n}_{\gamma_1} u^{n}_{\gamma_2}\mbox{~~for $j=n$}
\end{array}\right.
\label{ibejejgeneral}
\ee
\be
\begin{array}{c}
\Braket{0|{\rm T}\,
\phi_{\alpha_1\cdots \alpha_\mu;\beta_1\cdots \beta_\nu}
(x)\,
\phi^*_{\alpha_1\cdots \alpha_\mu;\beta_1\cdots \beta_\nu}
(0)|0}
\\
\displaystyle
=\biggl(S_1(x)\,
S_2(x)\biggr)^{\mathbf{N}(A)+\mathbf{N}(B)}
\exp
\left[\rule{0ex}{6ex}\right.
2\kappa(m_1,x)\left(\left(\sum_{\gamma =1}^2(N^A_\gamma +N^B_\gamma )\right)
-\frac{1}{2}\left(\sum_{\gamma =1}^2(N^A_\gamma -N^B_\gamma )\right)^2\right)
\\
\displaystyle
+\sum_{j=2}^{n-1}
2\kappa(m_j,x)\left(N^{AB}_{j+1}+N^B_{j+1}
-\frac{1}{j+1}\left(\sum_{\gamma =1}^{j+1}(N^A_\gamma -N^B_\gamma )\right)^2
+\frac{1}{j}\left(\sum_{\gamma =1}^{j}(N^A_\gamma -N^B_\gamma )\right)^2\right)	
\\
\displaystyle
+2\kappa(m_n,x)
\,\frac{1}{n}\left(\sum_{\gamma =1}^{n}(N^A_\gamma -N^B_\gamma )\right)^2
\left.\rule{0ex}{6ex}\right]
\end{array}
\label{ibmjexponent}
\ee
where $m_n\ll m_j$ for $j<n$.  This has no effect at all on the ZDOP
correlators.  It has very little effect on the 
LDOP correlators for $x\ll1/m_n$
{\bf But at very large distances the last term in (\ref{ibmjexponent}) cancels
the LDOP anomalous dimensions and all the LDOPs become ZDOPs, the conformal
sector completely disappears and the
theory has a mass gap, which we interpret to mean
that in this modified version
the fermions are completely bound into massive 
particles, albeit at very large distance.}  

Because
there is a mass gap in the theory with the $U(1)$ gauge interaction
there is no unparticle stuff.  
All the scalar unparticle fields disappear from the low-energy
physics by becoming ZDOPs.  The baryonic unparticle operators are absent
for a different reason --- they are not $U(1)$ gauge
invariant.  Presumably, baryon-antibaryon states are bound into massive
particles by the $U(1)$ gauge interaction.

{\figsize\begin{figure}[htb]
$$\includegraphics[height=.6\hsize]{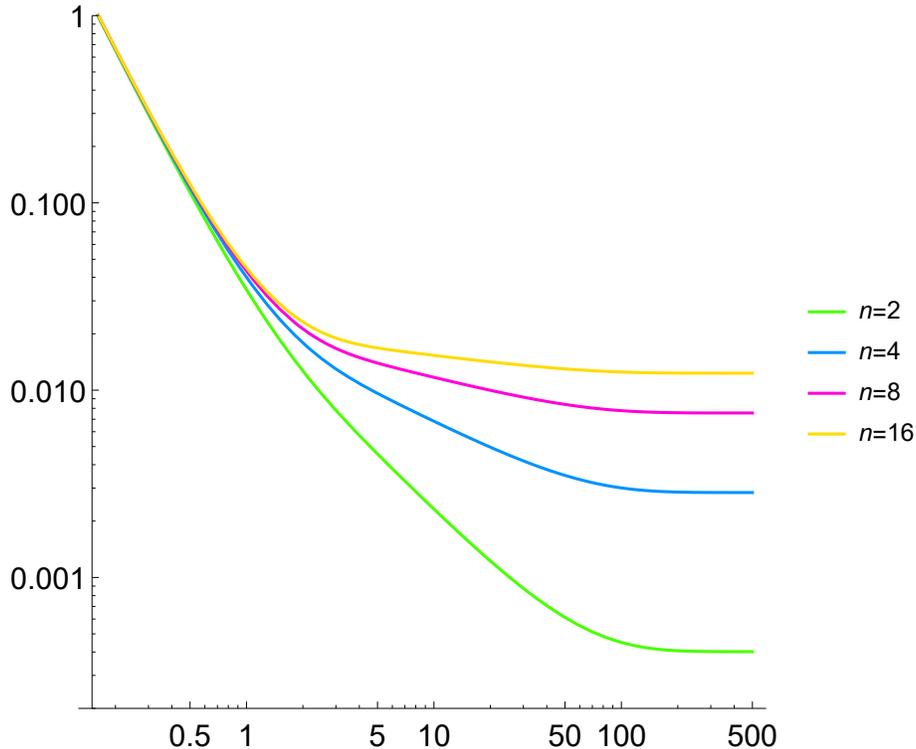}$$
\caption{\figsize\sf\label{fig-7}The LDOP correlation
function of figure~\ref{fig-2} for
$\mathbf{N}(A)=1$, $\mathbf{N}(B)=0$ with the addition of a $U(1)$ gauge
boson with dynamical mass $m/50$ as a function of spacelike
$x$ plotted in units of $1/m$ for various values of $n$, showing the
transition from free-fermion 
behavior at small $x$ to unparticle behavior for $1/m\ll x\ll 50/m$ to
constancy for $x\gg50/m$.}\end{figure}}
The correlators, in this case, look very different from the ZDOP correlators
in figures~\ref{fig-4}-\ref{fig-6}.   In figure~\ref{fig-7}, we
illustrate this by showing what happens to the LDOP correlation
function of figure~\ref{fig-2} for
$\mathbf{N}(A)=1$, $\mathbf{N}(B)=0$ when we add to the Lagrangian 
a $U(1)$ gauge
boson with dynamical mass $m/50$.
In this figure, you can see the
transition from free-fermion 
behavior at small $x$ to unparticle behavior for $1/m\ll x\ll 50/m$ to
constancy for $x\gg50/m$.  In the unparticle regime,  $1/m\ll x\ll 50/m$,
the free-fermion contribution is completely canceled by the ghost
contribution, leaving only the effect of the anomalous dimension.  In ZDOP
correlators in the $SU(n)$ theory, there is always a free fermion
contribution until the correlator becomes constant.\footnote{Note that
figure~\ref{fig-2} and figure~\ref{fig-7} have the same aspect ratio, while
to make figures~\ref{fig-4}-\ref{fig-6} fit, we have different aspect
ratios and this 
somewhat obscures the effect we are describing.} 

The light scalar with mass $m_n$ 
is the partner of the ghost that cancels the anomalous
dimensions of the LDOPs.   It is a pseudo-scalar $\eta'$ associated with the
dynamical breaking of the chiral $U(1)$.  

\section{Conclusions\label{sec-end}}

We hope that we have convinced our readers that it is interesting to study
generalized Schwinger models with
long-distance zero-dimension operators as well as unparticle stuff.  
While many of the properties of these
models are 
dependent on the special properties of massless fermions in 1+1 dimensions,
some of the 
phenomena we observe here such as conformal coalescence could occur in
more interesting
theories. We look forward to further exploration of this class of
models.

We believe that it might be interesting to simulate the model of
section~\ref{sec-ib} on the lattice (for diagonal color and/or other
couplings satisfying (\ref{eoverm})).  If one starts with $m_j=m_n\,\forall\,
j$, the model should be equivalent to $n$ copies of the Schwinger model.
Then one could try to lower $m_n$ with $m_j$ fixed 
and study the physics between $m_n$ and
$m_j$ for $j<n$.  This might provide an
interesting window into unparticle physics on the lattice. 

Another direction from section~\ref{sec-ib} is to consider the opposite limit,
$m_n\gg m_j$ for $j<n$.  This is a version of the $n$-flavor Schwinger
model modified in the IR to give complete binding and a mass gap.  It would
be interesting to study the connection of this limit with the $n$-flavor
massive Schwinger model.~\cite{Coleman:1976uz}

We further hope that the picture described in section~\ref{sec-ib} of
unparticle behavior as incomplete binding may be useful in 3+1 dimensions.

\section*{Acknowledgments\label{sec-ack}}

This work is supported in part by NSF grant
PHY-1719924.  BN's research is supported in part by the Harvard College
Research Program.

\bibliography{up4}

\appendix

\section{ZDOP VEV with multiple masses\label{sec-vevmasses}}

If $\mathbf{N}(A)=\mathbf{N}(B)$ so that $\phi_{A;B}$ is a ZDOP,
(\ref{zdopvev}) goes to
\be
\begin{array}{c}
\displaystyle
\rule{0ex}{6ex}
\Braket{0|
\phi_{A;B}
|0}=e^{i\theta_{A;B}}
\left(\frac{\xi}{2\pi}\right)^{2\mathbf{N}(A)}
\exp\left[\sum_{j=1}^{n-1}
\left(\sum_{\gamma_1\in A }-\sum_{\gamma_1\in B }\right)
\left(\sum_{\gamma_2\in A }-\sum_{\gamma_2\in B }\right)
\frac{e_{j{\gamma_1}}e_{j{\gamma_2}}}{\pi m_j}
\log(m_j)\right]
\end{array}
\label{gadmjz2}
\ee
\be
\begin{array}{c}
\displaystyle
=e^{i\theta_{A;B}}
\left(\frac{\xi}{2\pi}\right)^{2\mathbf{N}(A)}
\exp
\left[\rule{0ex}{6ex}\right.
\log(m_1)\left(\left(\sum_{\gamma =1}^2(N^A_\gamma 
+N^B_\gamma )\right)
-\frac{1}{2}\left(\sum_{\gamma =1}^2(N^A_\gamma 
-N^B_\gamma )\right)^2\right)
\\
\displaystyle
+\sum_{j=2}^{n-1}
\log(m_j)\left(N^{A}_{j+1}+N^B_{j+1}
-\frac{1}{j+1}\left(\sum_{\gamma =1}^{j+1}(N^A_\gamma 
-N^B_\gamma )\right)^2
+\frac{1}{j}\left(\sum_{\gamma =1}^{j}(N^A_\gamma 
-N^B_\gamma )\right)^2\right)	
\left.\rule{0ex}{6ex}\right]
\end{array}
\label{mjexponentz}
\ee
Where $N^C_\gamma\equiv \mathbf{N}\bigl(C\cap\{\gamma\}\bigr)$ is
defined in (\ref{omega}).
If 
\be
N^A_\gamma=\delta_{\gamma j}
\mbox{~and~}
N^B_\gamma=\delta_{\gamma }
\mbox{~for~}2<j< k,
%\label{}
\ee
This simplifies to
\be
\Braket{0|
\phi_{A;B}
|0}=e^{i\theta_{A;B}}
\left(\frac{\xi}{2\pi}\right)^{2}
\exp\left[\sum _{l=j-2}^{k-2} \frac{\log (m_{l})}{l (l+1)}+\frac{(j-2) \log
   (m_{j-1})}{j-1}+\frac{k \log (m_{k-1})}{k-1}\right]
\label{vevmasses}
\ee
Depending on all the masses from $m_{j-1}$ to $m_{k-1}$

\section{Signs\label{sec-signs}}
Consider the 3-point correlators of the form of (\ref{phi3}).  In these
correlators there may be signs that come from the Fermi statistics of the
free-fermion factors in (\ref{psi-alpha}).  The bosonic fields contribute
exponential factors that we calculate exactly.  Thus in determining the
signs of the correlators for spacelike separations, we can treat the
fermions as free. If 
$A_1$ and $B_2$ have an index $\alpha$ in common, then we claim that the
sign of (\ref{phi3}) is the same as
the sign of the reduced correlator 
\be
\Braket{0|{\rm T}\,
\phi_{A_1-\alpha;B_1}
(x_1)\,
\phi_{A_2;B_2-\alpha}
(x_2)\,
\phi_{A_3;B_3}
(x_3)
|0}
\label{phi3r}
\ee
with the shared index removed.  To see this, we will write the free-fermion
part of the correlator as
\be
\Braket{0|{\rm T}\,
\bar\chi_{1A_1}(x_1)\bar\chi_{2B_1}(x_1)
\chi_{1\tilde B_1}(x_1)\chi_{2\tilde A_1}(x_1)\,
\bar\chi_{1A_2}(x_2)\bar\chi_{2B_2}(x_2)
\chi_{1\tilde B_2}(x_2)\chi_{2\tilde A_2}(x_2)\,
\phi_{A_3;B_3}
(x_3)
|0}
\label{phichi}
\ee
where
\be
\bar\chi_{jC}(x)=\prod_{\alpha\in C}\Psi^*_{j\alpha}(x)
\quad\quad
\chi_{j\tilde C}(x)=\prod_{\alpha\in \tilde C}\Psi_{j\alpha}(x)
\label{chi}
\ee
where the $\tilde C$ represents the same indices as in $C$ but in the
opposite order.
Next we will use our freedom to
reorder the indices in the sets $A_1$ and $B_2$ so that $\alpha$ is the
last index in $A_1$ and the first index in $B_2$.  Then we can write
(\ref{phichi}) as 
\be
\begin{array}{c}
\bigl\langle 0\,\bigr|{\rm T}\,
\displaystyle
\bar\chi_{1(A_1-\alpha)}(x_1)\Psi^*_{1\alpha}(x_1)\bar\chi_{2B_1}(x_1)
\chi_{1\tilde B_1}(x_1)\Psi_{2\alpha}(x_1)\chi_{2(\tilde A_1-\alpha)}(x_1)\\
\displaystyle
\bar\chi_{1A_2}(x_2)\Psi^*_{2\alpha}(x_2)\bar\chi_{2(B_2-\alpha)}(x_2)
\chi_{1(\tilde B_2-\alpha)}(x_2)\Psi_{1\alpha}(x_2)\chi_{2\tilde A_2}(x_2)\,
\phi_{A_3;B_3}
(x_3)
\bigl|\,0\bigr\rangle
\end{array}
\label{phichi2}
\ee
and we can write this as
\be
\begin{array}{c}
\bigl\langle 0\,\bigr|{\rm T}\,
\displaystyle
\bar\chi_{1(A_1-\alpha)}(x_1)\Psi^*_{1\alpha}(x_1)\Psi_{1\alpha}(x_2)
\bar\chi_{2B_1}(x_1)
\chi_{1\tilde B_1}(x_1)\chi_{2(\tilde A_1-\alpha)}(x_1)\\
\displaystyle
\Psi_{2\alpha}(x_1)\Psi^*_{2\alpha}(x_2)
\bar\chi_{1A_2}(x_2)\bar\chi_{2(B_2-\alpha)}(x_2)
\chi_{1(\tilde B_2-\alpha)}(x_2)\chi_{2\tilde A_2}(x_2)\,
\phi_{A_3;B_3}
(x_3)
\bigl|\,0\bigr\rangle
\end{array}
\label{phichi3}
\ee
picking up an even number of minus signs
In the 3-point function
(\ref{phi3}), all of the contractions of the free-fermion fields are unique
and so (\ref{phichi3}) is
\be
\Braket{0|{\rm T}\,\Psi^*_{1\alpha}(x_1)\Psi_{1\alpha}(x_2)
|0}
\Braket{0|{\rm T}\,\Psi_{2\alpha}(x_1)\Psi^*_{2\alpha}(x_2)
|0}
=\frac{1}{4\pi^2\bigl(-(x_1-x_2)^2+i\epsilon\bigr)}
\label{phichi4}
\ee
which is positive for spacelike separation.  

The $\phi$s are bosonic operators so there 
is nothing special about what we have called 1 and 2 in this
argument.  
Thus we can
eliminate any pair of common indices from $A_j$ and $B_k$
(where $j\neq k$ because of (\ref{acapb})) 
with no change in sign.
We can then continue eliminating pairs of indices until we have eliminated
all the fermions and our reduced correlator is $1$ so we have a a positive
sign for our correlator at spacelike separation..  

In the $r$-point function (\ref{phir}) for $r>3$ we have to consider the possibility of
repeated indices.  Suppose that an index $\alpha$ is repeated $\ell$ times in
the $A$s.  (\ref{uaub}) implies that it is also repeated $\ell$ times in
the $B$s  and (\ref{acapb}) implies that all the $\alpha$ indices are in
different $\phi$s.  Our labels are arbitrary, so we can assume without loss of
generality that the $\alpha$ label appears in the sets
\be
A_1,\,B_2,\cdots A_{2\ell-1},B_{2\ell}
%\label{}
\ee
Then we repeat the analysis of (\ref{phichi})-(\ref{phichi4}) and conclude
that the free-fermion part of (\ref{phir}) is
\be
\Braket{0|{\rm T}\,\prod_{k-1}^\ell \Psi^*_{1\alpha}(x_{2k-1})\Psi_{1\alpha}(x_{2k})
|0}
\Braket{0|{\rm T}\,\prod_{k-1}^\ell \Psi_{2\alpha}(x_{2k-1})\Psi^*_{2\alpha}(x_{2k})
|0}
\label{phichir}
\ee
times the free-fermion part of the reduced correlator with the $\alpha$
indices missing.  Then using (\ref{magic}) we can write (\ref{phichir})
as
\be
\left(\frac{1}{4\pi^2}\right)^\ell\frac{\displaystyle\prod_{j<k=1}^\ell \left(
-(x_{2j-1}-x_{2k-1})^2+i\epsilon\right)\left(- (x_{2j}-x_{2k})^2+i\epsilon\right)
}{\displaystyle\prod_{j,k=1}^{\ell} \left(- (x_{2j-1}-x_{2k})^2+i\epsilon\right)}
\label{magicr}
\ee
which is positive for spacelike separations.
Then just as for $r=3$ we can eliminate all the indices and conclude that
(\ref{phir}) is positive for spacelike separations.

\end{document}